\documentclass[12pt,superscriptaddress,showpacs,showkeys,twocolumn,prb,aps,reprint,a4paper,floatfix]{revtex4-1}

\usepackage{graphicx}
\usepackage{amsmath, amsthm, amssymb, amsfonts}
\usepackage{float}

\usepackage{natbib}
\usepackage{physics}

\usepackage{array}
\usepackage{ragged2e} 

\makeatletter
\newcommand*{\citenst}[2][]{%
  \begingroup
  \let\NAT@mbox=\mbox
  \let\@cite\NAT@citenum
  \let\NAT@space\NAT@spacechar
  \let\NAT@super@kern\relax
  \renewcommand\NAT@open{[}%
  \renewcommand\NAT@close{]}%
  \citet[#1]{#2}%
  \endgroup
}
\makeatother

\makeatletter
\newcommand*{\citenumns}[2][]{%
  \begingroup
  \let\NAT@mbox=\mbox
  \let\@cite\NAT@citenum
  \let\NAT@space\NAT@spacechar
  \let\NAT@super@kern\relax
  \renewcommand\NAT@open{[}
  \renewcommand\NAT@close{]}%
  \cite[#1]{#2}
  \endgroup
}
\makeatother

\begin{document}
\title{Quantum analysis of second-order effects in superconducting travelling-wave parametric amplifiers}
\author{Songyuan Zhao}
\email{sz311@cam.ac.uk}
\author{Stafford Withington}
\date{\today}

\affiliation{Cavendish Laboratory, JJ Thomson Avenue, Cambridge CB3 OHE, United Kingdom.}

\begin{abstract}
\noindent
We have performed a quantum mechanical analysis of travelling-wave parametric amplifiers (TWPAs) in order to investigate five experimental phenomena related to their operations, namely the effect of impedance mismatch, the presence of upper idler modes, the presence of quantum and thermal noise, the generation of squeezed states, and the preservation of pre-squeezed states during amplification. Our analysis uses momentum operators to describe the spatial evolution of quantised modes along a TWPA. We calculate the restriction placed on pump amplitude as well as amplifier gain as a result of impedance mismatch between a TWPA and its external system. We apply our analysis to upper idler modes and demonstrate that they will result in suppressed gain. We show that an ideal TWPA is indeed quantum-limited - i.e. it introduces a half-quantum of zero-point fluctuation which is the minimum possible noise contribution for a phase-preserving linear amplifier. We analyse the thermal noise associated with a TWPA by considering the effect of distributed sources along an amplifier transmission line. Our analysis predicts a doubling of thermal noise in the high gain limit as a result of wave-mixing between signal and idler modes. We study the operation of a TWPA in the presence of a DC bias current, and have shown that highly squeezed states can in principle be generated. However, amplifying a pre-squeezed state using a non-degenerate TWPA generally reduces the squeezing advantage.
\end{abstract}

\maketitle

\section{Introduction}

Travelling-wave parametric amplifiers (TWPAs) and  frequency convertors based on the nonlinear inductance of superconducting films have garnered significant theoretical \citenumns{Chaudhuri_2015,Erickson_2017,Songyuan_2019_paramp, songyuan2019_nonlinear} and experimental \citenumns{Eom_2012,Shan_2016,Adamyan_2016,Chaudhuri_2017,shu2021nonlinearity} interest. A variety of devices have demonstrated bandwidths of the order of a few gigahertz, critical currents of the order of a few milliamperes, and noise temperatures close to the standard quantum limit (SQL) \citenumns{Eom_2012,Bockstiegel_2014,Adamyan_2016,Vissers_2016,Chaudhuri_2017}. These characteristics make TWPAs well suited to applications that require broadband, low noise amplification at microwave frequencies, such as qubit and detector readout \citenumns{Ranzani_2018, Zobrist_2019, Zorin_2019, Vissers_2020}. For example, the high fidelity readout of qubits is a requirement of many quantum information tasks such as error correction \citenumns{Barends_2014}, quantum feedback control \citenumns{Vijay_2012}, and the implementation of complex algorithms \citenumns{Riste_2017}. Likewise, low-noise receivers are essential for many astronomical observations \citenumns{Monfardini_2010,Maloney_2010, Endo_2012, Mazin_2013} and fundamental physics experiments, such as neutrinoless double-beta decay experiments \citenumns{Battistelli2015, Cardani_2015}, direct measurement of neutrino mass \citenumns{Project8_2009,Hao_2019,Saakyan_2020}, and dark matter searches \citenumns{Golwala2008, Cornell_2018}.

Despite the many recent successes in designing and fabricating superconducting  TWPAs, a number of unexplored issues remain. These have important practical implications, but have not, to our knowledge, been dealt with at a detailed theoretical level. In particular, we are concerned about (i) the effect of reflections from the ends of TWPA chips, and connectors, giving rise to resonant longitudinal modes, which reveal themselves through rapid gain variations with frequency. Such variations are seen in most, if not all, physical realisations, but their origin is uncertain. (ii) In simulations to date, the dynamical equations are solved by ignoring certain frequency-conversion terms, essentially making a rotating wave approximation, but the effect of these terms is to change the functional form of gain with propagation length. Again we are concerned about the effects of these terms on amplifier performance. (iii) TWPAs are fabricated from materials having high normal-state resistivity, because this should enhance the kinetic inductance effect, but then a question arises about the growth of thermal noise with propagation distance, and the implications this has for achieving the SQL with modest bath temperatures: say 1 K as distinct from 10 mK. (iv) After having shown that the TWPA is capable of achieving the SQL, we then study whether such a component can be used in applications requiring squeezing. We are not aware of superconducting TWPA being used in this way, and so discuss whether a superconducting TWPA can be operated in a degenerate mode to produce a squeezed state of microwave radiation. (v) Finally, we study the spatial evolution of a squeezed state through a non-degenerate parametric amplifier, and the degree to which the original squeezing is preserved.

To solve (i), we must consider reflections, and the reverse propagation of a signal along a TWPA. We demonstrate that, due to the reflected pump mode, the signal mode can have appreciable gain in the backward direction. A reflected signal on its own has little effect, but a reflected pump and signal can lead to an appreciable loop gain, which in the extreme case may lead to instability. When the loop gain is smaller than unity, reflections introduce ripples into gain plots; and when the loop gain is greater than unity, reflections may result in instability. We compute the restriction the loop gain places on the choice of pump amplitude and the maximum forward gain. From a practical perspective, special efforts should be made to eliminate impedance mismatch at the ends of TWPA lines in order to reduce gain ripples as well as to maximise forward gain.

To solve (ii), we must consider the effects of upper idler modes, which exist above twice the pump frequency. Using the quantum framework presented here, we show that these modes periodically siphon energy away from the signal modes, and as a result suppress gain in a spatially periodic way.

To understand (iii), we apply our framework to the roles of quantum and thermal noise. Whereas the quantum noise of a TWPA is intrinsically related to the preservation of phase during amplification, the thermal noise arises from the non-zero microwave resistivity of the superconducting transmission line. Theoretical studies \citenumns{Caves_1982} on linear amplifiers have demonstrated that the quantum noise is constrained by the fundamental theorem
\begin{align}
  A\geq \frac{1}{2}\left|1-g^{-1}\right| \,,
\end{align}
where $A$ is the added noise, in terms of number of quanta, and $g$ is the amplitude gain. In the case of high gain, and when the equality applies, $A=1/2$, which is often referred to as the standard quantum limit: a half-quantum zero-point fluctuation. Experimental realizations have shown that TWPAs can indeed achieve additive noise approaching the SQL \citenumns{Eom_2012}. (If the signal is sufficiently broadband such that the idler also contributes to the signal, it is possible to achieve parametric amplification beyond the SQL \citenumns{Renger_2020}. Our present paper is focused on the conventional operation of a TWPA where there is no overlap between the signal and the idler frequencies, and thus the relevant amplification noise limit is the SQL.) In this study, we show, by analysing quantum noise in a transmission line having a distributed non-linear inductance, that ideal superconducting TWPAs are governed by the equality, thereby providing a theoretical explanation of reported observations, and indicating that the SQL should in principle be achievable. However, in addition to the intrinsic quantum noise, thermal noise will be present, and this determines the temperature to which a TWPA must be cooled to achieve the SQL. The conflict is that kinetic inductance nonlinearity increases with resistivity \citenumns{Eom_2012,songyuan2019_nonlinear}, and so high resistivity materials are beneficial, but thermal noise is also likely to increase with resistivity \citenumns{Johnson_1928}. Without loss of generality, we ignore dielectric loss, because we are most interested in the relationship between nonlinear kinetic inductance and noise. To analyse the effect of thermal noise, we investigate the general evolution of wave modes in the presence of distributed sources, and then present a mathematical formalism that captures the effect of the noise on overall amplifier performance. Our analysis predicts a doubling of thermal noise in the high gain limit as a result of wave-mixing between the signal and idler modes. This fundamental effect should be borne in mind when designing superconducting TWPAs.

To study these important effects, we need to quantise the field on the transmission line. Historically, the first travelling-wave parametric amplifiers were demonstrated at optical wavelengths \citenumns{Giordmaine_OPA_1965}. These devices make use of field dependent nonlinear susceptibilities, and have been studied extensively in the context of nonlinear optics \citenumns{Manley_Rowe_1956,Hansryd_2002,Kylemark_2004}. A quantum theory of parametric amplification was presented by Louisell, Yariv, and Siegman in the 1960s \citenumns{Louisell_1961}, and is still in wide use today \citenumns{Clerk_2010,Eom_2012}. The theory is focused on cavity parametric amplifiers, and as a result is based on a Hamiltonian approach, where longitudinal resonant modes are quantised. Problems arise, however, when the theory is applied to TWPAs, because as noted by other authors, the Hamiltonian formalism is unsuitable for describing phenomena that involve propagation in dispersive linear media \citenumns{Abram_1987,Hutter_1990}, let alone the non-linear medium considered here. In the case of TWPAs based on non-linear superconducting films, the total Hamiltonian cannot be calculated straightforwardly from the Hamiltonian density because the wave amplitudes vary with position. In this paper we adopt a different approach, and adapt the momentum operator formalism described in \citenumns{Hutter_1990} for the purpose of studying the spatial evolution of signal and noise in superconducting TWPAs.

\section{Principles of operation}
Our model of a TWPA consists of a superconducting transmission line with matched input and output terminations biased by a strong pump current
\begin{align}
  I_p = I_{p,0}\cos{\left(k_p z-\omega_p t + \Phi_p\right)} \,, \label{eq:pump_form}
\end{align}
where $I_{p,0}$ is the amplitude of the pump, $k_p$ is the pump wave vector, $\omega_p=k_p v$ is the pump angular frequency, $v$ is the phase velocity, and $\Phi_p$ is the pump phase which is considered to be a constant in this study. We assume that the effect of the nonlinear kinetic inductance can be captured by the second order expansion
\begin{equation}
  L = L_0[1+(I/I_*)^2] \, , \label{eq:nonlinear_scale}
\end{equation}
where $L_0$ is the inductance per unit length in the absence of the inductive nonlinearity, $I$ is the line current, and $I_*$ characterises the scale size of the parametric process \citenumns{Eom_2012,songyuan2019_nonlinear}. We also assume that the amplifier is operating in a regime where pump current is not depleted.

To analyse spatial evolution, we adopt a quantum field theoretic approach, and relegate position on the transmission line $z$ from a dynamic variable to a label, along with time $t$. Both current $\hat{I}(z,t)$ and potential difference $\hat{V}(z,t)$ are then quantum fields. We introduce the generalized flux field $\hat{\phi}(z,t)$ defined by
\begin{equation}
  \hat{\phi}(z,t) = \int^{t}_{-\infty} dt' \hat{V}(z,t')\,.
\end{equation}
It can be shown that
\begin{align}
  \hat{I}(z,t) = -\frac{1}{L}\frac{\partial\hat{\phi}}{\partial z}\,, \label{eq:current}\\
  \hat{V}(z,t) = \frac{\partial\hat{\phi}}{\partial t}\,.
\end{align}

We expand the flux field in terms of quantised modes
\begin{equation}
  \hat{\phi}(z,t) = \sqrt{\frac{Z_0}{2 \tau}}\sum_m\sqrt{\frac{\hbar}{\omega_m}}\left[\hat{b}_m(z,t)+\hat{b}_m^\dagger(z,t)\right]\,, \label{eq:modes_expansion}
\end{equation}
where $Z_0$ is the transmission line characteristic impedance, $\tau$ is the time of travel along the transmission line, and is much larger than any period of relevant frequencies, $\hbar$ is the reduced Planck constant, $\hat{b}_m(z,t)=\hat{a}_m(z) \exp{i(k_m z-\omega_m t)}$, and $\hat{a}_m(z)$ is the annihilation operator of the m-th frequency mode which varies slowly with position $z$ (compared to the rapidly varying $\exp{i(k_m z-\omega_m t)}$ term). For brevity, we shall often drop explicit reference to $z$, where appropriate: $a_{m} \equiv a_{m}(z)$. $\hat{a}_m^\dagger$ denotes the Hermitian adjoint of $\hat{a}_m$, and is the creation operator of the m-th frequency mode.

The equal-space commutation relation of the mode operators \citenumns{Hutter_1990} is given by
\begin{align}
  \left[\hat{a}_{m},\hat{a}_{n}^\dagger\right]=\delta_{m,n}\,,
\end{align}
where $\delta_{m,n}$ is the Dirac delta function, and the square bracket denotes commutation, i.e. $\left[\hat{a}_{m},\hat{a}_{n}^\dagger\right]=\hat{a}_{m}\hat{a}_{n}^\dagger-\hat{a}_{n}^\dagger\hat{a}_{m}$.

Following \citenumns{Hutter_1990}, we propagate operators using the momentum counterpart to the Heisenberg picture, such that all spatial dependence is in the operators. The momentum picture is relevant to travelling-wave devices because its equation of motion evolves appropriate operators in space, whereas the equation of motion in the Heisenberg picture evolves operators in time. Further, since $\hat{a}_m$ is spatially-varying in travelling-wave devices, it is easier computing the total momentum from the momentum \textit{flux} than computing the total Hamiltonian from the Hamiltonian \textit{density}. In Appendix A, we demonstrate that a quantum version of the Telegrapher's equations can be readily derived from the momentum operator formalism. The mode operator in the momentum picture is $\hat{c}_m=\hat{a}_m \exp(ik_m z)$. In order to evolve the quantised modes, we employ the equation of motion in the momentum picture
\begin{align}
  i\hbar \frac{d\hat{A}_M}{dz}=\left[\hat{A}_M,\hat{G}_M\right]+i\hbar \frac{\partial\hat{A}_S}{\partial z} \,,
\end{align}
where $\hat{G}$ is the momentum operator, and $\hat{A}$ is any general operator. The Schrodinger picture operator is denoted by subscript $S$, whereas momentum picture operators are denoted by subscript $M$. $\hat{G}$ can be calculated by integrating the momentum flux $\hat{\mathcal{G}}$
\begin{align}
  \hat{G}=\int_{t_0}^{t_0+\tau}dt\,\hat{\mathcal{G}} \,. \label{eq:integrate_momentum}
\end{align}

The operators in this study do not have any intrinsic spatial dependence, beyond that associated with the ordinary dynamical behaviour, and so we simplify the equation as follow:
\begin{align}
  i\hbar \frac{d\hat{A}_M}{dz}=\left[\hat{A}_M,\hat{G}_M\right] \,. \label{eq:EoM_Heisenberg}
\end{align}

In this section, we restrict the frequencies present to the signal $\omega_s$, the pump $\omega_p$, and the idler $\omega_i=2\omega_p-\omega_s$. We will relax this assumption later. We now divide the momentum operator into the linear part $\hat{G}_0$ and the nonlinear part $\hat{G}_{4WM}$. For a linear transmission line
\begin{align}
  \hat{\mathcal{G}}(z,t)=\frac{1}{2}\left(C\hat{V}(z,t)^2+L\hat{I}(z,t)^2\right) \,, \\
  \hat{G}_0=\sum_m \hbar k_m \left(\hat{a}_m^\dagger\hat{a}_m+\frac{1}{2}\right)\,,
\end{align}
where $C$ is the capacitance per unit length.

We introduce nonlinearity by substituting equation~(\ref{eq:pump_form}) into equation~(\ref{eq:nonlinear_scale}). The extra term due to pump modulated kinetic inductance now results in extra terms in the momentum operator. We adopt the same reasoning presented in \citenumns{Louisell_1961}, which states that non-energy-conserving terms will quickly average to zero in the long integration of equation~(\ref{eq:integrate_momentum}). Keeping only the energy conserving term, the extra terms in the momentum operator are
\begin{align}
  \hat{G}_{4WM}=-\hbar C_{i,s}\left[\exp(i2\Phi_p)\hat{a}_i^\dagger\hat{a}_s^\dagger\right.\,\notag \\
  \left.+\exp(-i2\Phi_p)\hat{a}_i\hat{a}_s\right]\,,
\end{align}
where
\begin{align}
  C_{i,s}=\frac{1}{8}\frac{L_0}{L_1}\left(\frac{I_{p,0}}{I_*}\right)^2\frac{\sqrt{\omega_i\omega_s}}{v}\,,
\end{align}
and $L_1=L_0[1+(I_{p,0}/I_*)^2/2]\approx L_0$.
Applying the equation of motion to mode operators in the momentum picture, and then translating the results back to the interaction picture, we obtain the following system of equations
\begin{align}
  \frac{\partial}{\partial z}\hat{a}_s=-iC_{i,s}\exp(i2\Phi_p)\hat{a}_i^\dagger \, \\
  \frac{\partial}{\partial z}\hat{a}_i^\dagger=iC_{i,s}\exp(-i2\Phi_p)\hat{a}_s^\dagger \, ,
\end{align}
which has solution
\begin{align}
  \hat{a}_s=&\hat{a}_{s,0}\cosh(C_{i,s}z)\notag \\
  &-i\exp(i2\Phi_p)\hat{a}_{i,0}^\dagger\sinh(C_{i,s}z) \, \label{eq:TWPA_soln_1} \\
  \hat{a}_i^\dagger=&\hat{a}_{i,0}^\dagger\cosh(C_{i,s}z) \notag \\
  &+i\exp(-i2\Phi_p)\hat{a}_{s,0}\sinh(C_{i,s}z) \, , \label{eq:TWPA_soln_2}
\end{align}
where $\hat{a}_{s,0}$ and $\hat{a}_{i,0}^\dagger$ correspond to the signal and idler modes at the input of the amplifier.

In practice, a parametric amplifier is often operated with the input state $\left|\mathrm{s}\right\rangle\left|\mathrm{i}\right\rangle=\left|a_{s,0}\right\rangle\left|0\right\rangle$, where $\left|\mathrm{s}\right\rangle$ and $\left|\mathrm{i}\right\rangle$ are the signal and idler states respectively, $\left|a_{s,0}\right\rangle$ is a coherent state with eigenvalue $a_{s,0}$, and $\left|0\right\rangle$ is the vacuum coherent state with eigenvalue $0$. In this case, hyperbolic amplification of the input signal is obtained:
\begin{align}
  \left\langle\hat{I}_{s}(z,t)\right\rangle &= \cosh(C_{i,s}z)\left\langle\hat{I}_{s,0}(z,t)\right\rangle=g\left\langle\hat{I}_{s,0}(z,t)\right\rangle\,, \\
  g &= \cosh(C_{i,s}z) \label{eq:amplitude_Gain}
\end{align}
where $g$ is the amplitude gain factor, and $\left\langle\hat{I}_{s,0}(z,t)\right\rangle$ is the expectation value of signal current at position $z$ and time $t$ in the absence of kinetic inductance nonlinearity (and thus without parametric amplification). This result is a well known feature of TWPAs \citenumns{Eom_2012,Chaudhuri_2015,Louisell_1961}, and indicates the validity of the approach. It is important to note that the gain here is independent of the pump phase $\Phi_p$. Consequently, the parametric gain is stable against phase fluctuations of the pump \citenumns{Songyuan_2019_paramp}. In the optical parametric amplifier community, phase modulation of the pump is sometimes intentionally introduced in order to mitigate back-scattering of the pump wave \citenumns{Hansryd_2002,Mussot_2004}.

\section{Backward propagation of reflected signal}

It is important to understand the behaviour of any reflected signal propagating against the strong forward pump current. This is because it is practically difficult to design a superconducting parametric amplifier where the input and output ports are perfectly impedance-matched to the external circuit. The phenomenon of gain ripple is often attributed to reflections \citenumns{Eom_2012,Bockstiegel_2014,Chaudhuri_2017,Adamyan_2016}, but in some cases the variations are quite extreme and occur with only very small changes in signal frequency.

There are a few constraints that limit the mixing possibilities between a strong \textit{forward}-propagating pump current and a weak \textit{backward}-propagating signal: (i) the interaction must involve at least one forward pump mode so as to have significant magnitude; (ii) the interaction must involve a signal mode in order to affect the state of the signal; (iii) the physics is governed by four-wave mixing; and (iv) in the absence of dispersion, the energy-momentum relation is constrained by $\hbar k =\pm \hbar\omega/v$, where the sign depends on the direction of propagation. Momentum and energy must be conserved simultaneously, and because the minus sign applies in the case of a backward travelling wave, the number of allowed mixing combinations is severely limited. In fact, because of the above restrictions, the only significant mixing process is the production of a pump-signal pair through the destruction of a pump-signal pair, which does not affect the amplitude of the signal, only the phase. Thus a backward traveling signal in the absence of a backward travelling pump, is neither parametrically attenuated nor amplified. It merely propagates as an ordinary wave on a linear transmission line, incurring ohmic and dielectric loss. This lack of a strong attenuation mechanism can be observed in measurements of the full two-port scattering parameters of a TWPA, such as figure~1 of reference \citenumns{Ranzani_2018}. This behaviour should be contrasted with that of say a microwave transistor amplifier, where a backward propagating signal is highly attenuated.

\begin{figure}[ht]
\includegraphics[width=8.6cm]{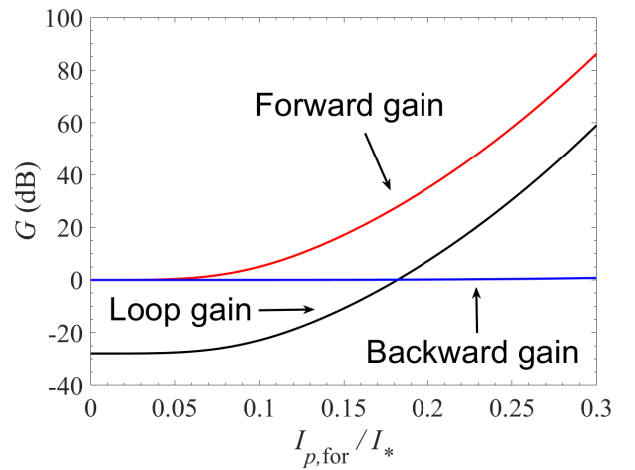}
\caption{\label{fig:75Ohm_Loop} Loop power gain, forward power gain, and backward power gain against forward pump strength for a $Z_{TWPA}=75\,\mathrm{\Omega}$ parametric amplifier with a $Z_{Ext} = 50\,\mathrm{\Omega}$ external impedance.}
\end{figure}

\begin{figure}[ht]
\includegraphics[width=8.6cm]{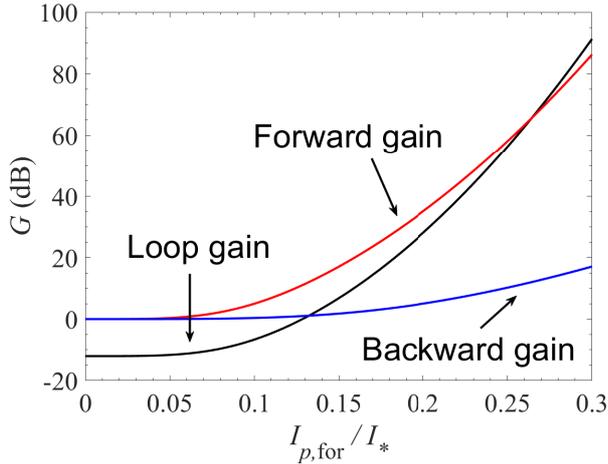}
\caption{\label{fig:150Ohm_Loop} Loop power gain, forward power gain, and backward power gain against forward pump strength for a $Z_{TWPA}=150\,\mathrm{\Omega}$ parametric amplifier with a $Z_{Ext} = 50\,\mathrm{\Omega}$ external impedance.}
\end{figure}

The second possible process concerns interactions between the \textit{reflected} signal and the \textit{reflected} pump, but this is the same as the forward parametric process, and so the effect on overall amplifier performance can be calculated using the analysis presented in section II. From equation~(\ref{eq:amplitude_Gain}), the gain is related to the pump current amplitude by
\begin{align}
  \mathrm{acosh}(g)\propto \left(I_{p,0}/I_*\right)^2\,,
\end{align} For the forward amplification process, $I_{p,0}$ takes on the value of the forward pump amplitude $I_{p,\mathrm{for}}$; for the backward amplification process, $I_{p,0}$ takes on the value of the backward reflected pump amplitude $I_{p,\mathrm{back}}$. A simple model can be constructed by relating $I_{p,\mathrm{for}}$ to $I_{p,\mathrm{back}}$ through $I_{p,\mathrm{back}}=\Gamma I_{p,\mathrm{for}}$, where $\Gamma = (Z_{TWPA}-Z_{Ext})/(Z_{TWPA}+Z_{Ext})$, $Z_{TWPA}$ is the TWPA characteristic impedance, and $Z_{Ext}$ is the characteristic impedance of the external system. From this, the loop gain of the signal mode can be calculated by including two reflections per loop, amplification by the forward pump, and amplification by the backward reflected pump. Here we have assumed that the TWPA is operated at pump power levels that avoid the onset of significant ohmic loss \citenumns{Eom_2012}. We have plotted the loop power gain, forward power gain, and backward power gain against forward pump strength $I_{p,\mathrm{for}}/I_*$ in figure~\ref{fig:75Ohm_Loop} for $Z_{TWPA}=75\,\mathrm{\Omega}$ and in figure~\ref{fig:150Ohm_Loop} for $Z_{TWPA}=150\,\mathrm{\Omega}$. $Z_{Ext}$ is taken to be $50\,\mathrm{\Omega}$ and $\mathrm{acosh}(g)/\left(I_{p,0}/I_*\right)^2$ is taken to be $120$, corresponding to a one meter long TWPA.

\begin{figure}[ht]
\includegraphics[width=8.6cm]{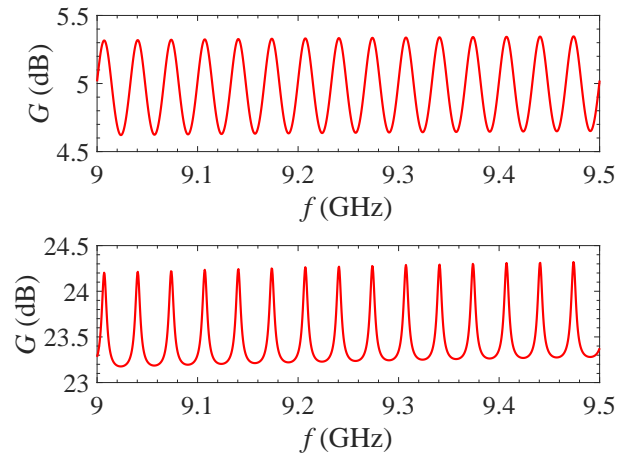}
\caption{\label{fig:Gain_ripple} Output gain $G$ against frequency $f$ for a $75\,\mathrm{\Omega}$ amplifier. Top figure: amplifier is operated with forward pump strength $I_{p,\mathrm{for}}/I_*=0.10$, which corresponds to loop gain of $-23.0\,\mathrm{dB}$; bottom figure: amplifier is operated with forward pump strength $I_{p,\mathrm{for}}/I_*=0.17$, which corresponds to loop gain of $-4.5\,\mathrm{dB}$..}
\end{figure}

As seen in the figures, both the forward and backward gain increase with applied pump power. The loop gain is particularly high for TWPA systems due to the lack of attenuation in the backward direction. The backward gain is close to unity when the impedance mismatch is small, but becomes significant when the mismatch is large. The loop gain at very small pump power arises as a consequence of having two reflections in the loop. As seen in figure~\ref{fig:150Ohm_Loop}, when the backward gain is greater than the reflection losses, the loop gain becomes greater than the forward gain. When the loop gain is larger than $0\,\mathrm{dB}$, at high input pump powers, the system may become unstable \citenumns{Carter_2018,Haidekker_2020}. When the loop gain is non-negligible but smaller than $0\,\mathrm{dB}$, the TWPA displays an unwanted interference pattern akin to a Fabry-Perot interferometer. Figure~\ref{fig:Gain_ripple} shows the gain profile of a $Z_{TWPA}=75\,\mathrm{\Omega}$, one meter long amplifier biased with a $10\,\mathrm{GHz}$ pump current. For the upper subfigure, the forward pump strength is given by $I_{p,\mathrm{for}}/I_*=0.10$, which corresponds to loop gain of $-23.0\,\mathrm{dB}$. For the lower subfigure, the forward pump strength is given by $I_{p,\mathrm{for}}/I_*=0.17$, which corresponds to loop gain of $-4.5\,\mathrm{dB}$. When the loop gain is much smaller than unity, the oscillation in frequency space is approximately sinusoidal. As the loop gain approaches unity from below, high-Q spectral features, which are typical of Fabry-Perot interferometers, appear in the gain profile. These high-Q features have been observed in various TWPA measurements, such as figure~3d of reference \citenumns{Eom_2012} and figure~4a of reference \citenumns{Adamyan_2016}.

\begin{figure}[ht]
\includegraphics[width=8.6cm]{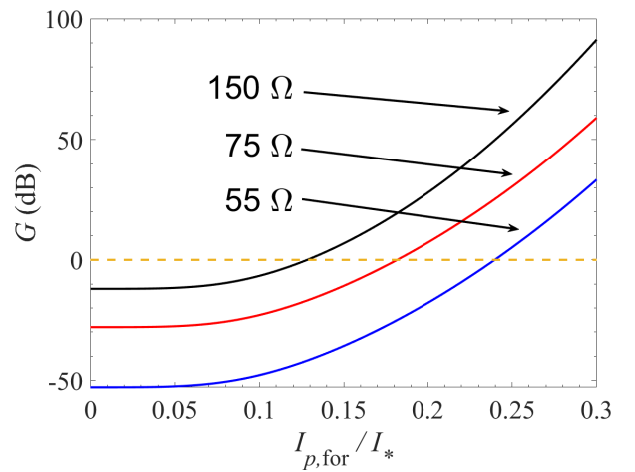}
\caption{\label{fig:vary_Z_loop} Loop power gain against forward pump strength for $Z_{TWPA}=150\,\mathrm{\Omega}$, $Z_{TWPA}=75\,\mathrm{\Omega}$, and $Z_{TWPA}=55\,\mathrm{\Omega}$ parametric amplifiers with a $Z_{Ext} = 50\,\mathrm{\Omega}$ external impedance.}
\end{figure}

\begin{figure}[ht]
\includegraphics[width=8.6cm]{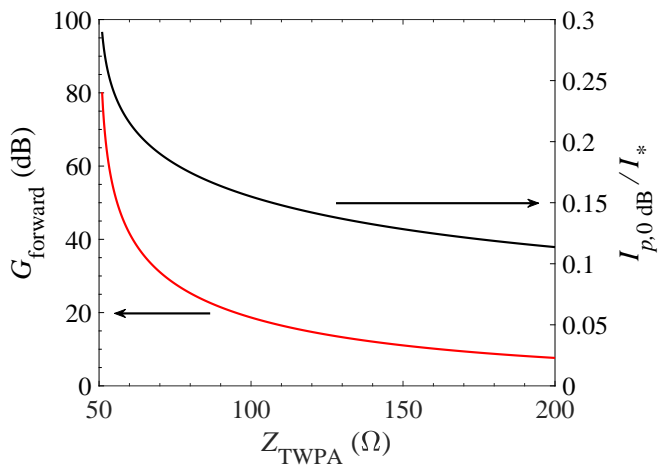}
\caption{\label{fig:max_Gain} Right axis, black plot: maximum current $I_{p,0\mathrm{dB}}/I_*$ such that the loop gain is $0\,\mathrm{dB}$ against device impedance $Z_{TWPA}$; left axis, red plot: maximum forward power gain $G_{\mathrm{forward}}$ such that the loop gain is $0\,\mathrm{dB}$ against device impedance $Z_{TWPA}$.}
\end{figure}

Figure~\ref{fig:vary_Z_loop} shows the loop gain for $Z_{TWPA}=150\,\mathrm{\Omega}$, $Z_{TWPA}=75\,\mathrm{\Omega}$, and $Z_{TWPA}=55\,\mathrm{\Omega}$, assuming a 50\,$\mathrm{\Omega}$ external system. The pump amplitude at which the loop gain is equal to $0\,\mathrm{dB}$ decreases as the impedance mismatch increases. We denote this pump level as $I_{p,0\mathrm{dB}}$. Since the characteristic impedance of a TWPA is fixed, for a given device, the pump current must be kept below $I_{p,0\mathrm{dB}}$ during operation to avoid instability. This requirement imposes a restriction on the maximum forward gain $G_{\mathrm{forward}}$. The dependence of both $G_{\mathrm{forward}}$ and $I_{p,0\mathrm{dB}}$ on $Z_{TWPA}$ are shown in figure~\ref{fig:max_Gain}. It can be seen that improving the impedance matching allows the TWPA to be operated at high pump current, achieving high gain.

%
%

 \section{Upper idler mode}

Thus far in our analysis, we have focused on the interplay between pump, signal, and idler modes. In practise, there are usually secondary effects arising from high-order intermodulation products. Experimentally, pump harmonics are often intentionally suppressed using impedance loading techniques \citenumns{Eom_2012}. Depending on the design, upper idler modes with frequencies $\omega_{u-s}=2\omega_{p}+\omega_{s}$ and $\omega_{u-i}=2\omega_{p}+\omega_{i}$ may inadvertently become significant. These upper idler modes do not require pump harmonics to be generated, and are thus important even in the presence of impedance loading. In cavity-based designs \citenumns{Louisell_1961}, upper idler modes can be safely ignored because amplifier cavities do not support strong resonances at upper idler frequencies. Transmission-line amplifiers, however, are naturally broadband devices, and so do not reject the upper idler modes. A few publications explicitly acknowledge the assumption that the upper idler modes are suppressed \citenumns{Jurkus_1960, Kononov_1989}, but none, to our knowledge, offer a description of their likely effect on performance.

In this section, we allow modes having the following frequencies to exist:
\begin{align}
   \omega_{p},\,\omega_{s},\,\omega_{i},\,\omega_{u-s},\,\omega_{u-i}\,.
\end{align}

Since we are loosening the restriction on the number of available frequencies, it is convenient to apply our method in the continuous frequency domain:
 \begin{align}
   \hat{I}(z,t)&=-i\sqrt{\frac{1}{2Z_0}}\int_0^\infty \frac{d\omega}{2\pi}\sqrt{\hbar\omega}\left[\hat{d}(\omega)-\hat{d}^\dagger(\omega)\right]\\
   \hat{V}(z,t)&=-i\sqrt{\frac{Z_0}{2}}\int_0^\infty \frac{d\omega}{2\pi}\sqrt{\hbar\omega}\left[\hat{d}(\omega)-\hat{d}^\dagger(\omega)\right] \label{eq:cont_V}\\
   \hat{d}(\omega)&=\hat{c}(\omega)\exp{i(kz-\omega t)}\,.
 \end{align}
 The equal-space commutation relation is then
 \begin{align}
   \left[\hat{c}(\omega,z),\hat{c}^\dagger(\omega',z)\right]=2\pi\delta(\omega-\omega')\,.
 \end{align}

The linear transmission line momentum operator $\hat{G}_0$ is
\begin{align}
  \hat{G}_0=\int \frac{d\omega}{2\pi}\frac{\hbar\omega}{v}\hat{c}^\dagger(\omega,z)\hat{c}(\omega,z)\,,
\end{align}
and the nonlinear momentum operator relevant to four-wave mixing, $\hat{G}_{4WM}$, is
 \begin{align}
  \hat{G}_{4WM}=& \hat{G}_1+ \hat{G}_2\,,\\
  \hat{G}_1=& -\hbar N \int \frac{d\omega_s}{2\pi}\sqrt{\omega_s\omega_i} \\
  &\left[e^{i2\Phi_p}\hat{c}^\dagger(\omega_s,z)\hat{c}^\dagger(\omega_i,z) \right. \notag \\
  &+\left.e^{-i2\Phi_p}\hat{c}(\omega_s,z)\hat{c}(\omega_i,z)\right]\, \notag \\
  \hat{G}_2=&\hbar N \int \frac{d\omega_s}{2\pi}\sqrt{\omega_s\omega_{u-s}} \\
  &\left[e^{i2\Phi_p}\hat{c}(\omega_s,z)\hat{c}^\dagger(\omega_{u-s},z)\right. \notag \\
  &+\left.e^{-i2\Phi_p}\hat{c}^\dagger(\omega_s,z)\hat{c}(\omega_{u-s},z)\right]\,. \notag
 \end{align}
$\hat{G}_1$ is the ordinary \textit{parametric amplifier} momentum operator, and $\hat{G}_2$ is a \textit{frequency-conversion} momentum operator, and $N=\left({I_{p,0}}/{I_*}\right)^2/\left({16v}\right)$.

Applying the equation of motion (\ref{eq:EoM_Heisenberg}), the overall momentum operator yields the following set of differential equations
 \begin{align}
   &\frac{d}{dz}\hat{c}(\omega_s,z)=-iA\hat{c}^\dagger(\omega_i,z)+iB\hat{c}(\omega_{u-s},z) \\
   &\frac{d}{dz}\hat{c}(\omega_i,z)=-iA\hat{c}^\dagger(\omega_s,z)+iC\hat{c}(\omega_{u-i},z) \\
   &\frac{d}{dz}\hat{c}(\omega_{u-s},z)=i\bar{B}\hat{c}(\omega_{s},z) \\
   &\frac{d}{dz}\hat{c}(\omega_{u-i},z)=i\bar{C}\hat{c}(\omega_{i},z)
 \end{align}
 where
 \begin{align}
   A=&2N\sqrt{\omega_s\omega_i}\exp(2\Phi_p) \notag \\
   B=&2N\sqrt{\omega_s\omega_{u-s}}\exp(-i2\Phi_p) \notag \\
   C=&2N\sqrt{\omega_i\omega_{u-i}}\exp(-i2\Phi_p) \,,\notag
 \end{align}
are constants, and the overbar denotes the complex conjugate. The terms associated with $A$ are the regular parametric amplifier terms; the terms associated with $B$ and $C$ are the frequency-conversion terms involving the upper idlers.

To understand the effect of having upper idler modes, we solved the above set of differential equations with respect to the amplifier input state $\left|\mathrm{s}\right\rangle \left|\mathrm{i}\right\rangle \left|\mathrm{u-s}\right\rangle \left|\mathrm{u-i}\right\rangle=\left|a_{s,0}\right\rangle \left|0\right\rangle \left|0\right\rangle \left|0\right\rangle$.
$\left|\mathrm{u-s}\right\rangle$ denotes the state of the upper idler at frequency $\omega_{u-s}$ and $\left|\mathrm{u-i}\right\rangle$ denotes the state of the upper idler at frequency $\omega_{u-i}$. For illustration purposes, we used the following parameters: inductance per unit length $L_0=8.1*10^{-7}\,\mathrm{Hm^{-1}}$, capacitance per unit length $C=2.8*10^{-10}\,\mathrm{Fm^{-1}}$, amplifier length $L_{PA}=1\,\mathrm{m}$, and pump frequency $f_{p}=10\,\mathrm{GHz}$. The simulations were carried out with $I_{p,0}/I_*=1/5$.

\begin{figure}[ht]
 \includegraphics[width=8.6cm]{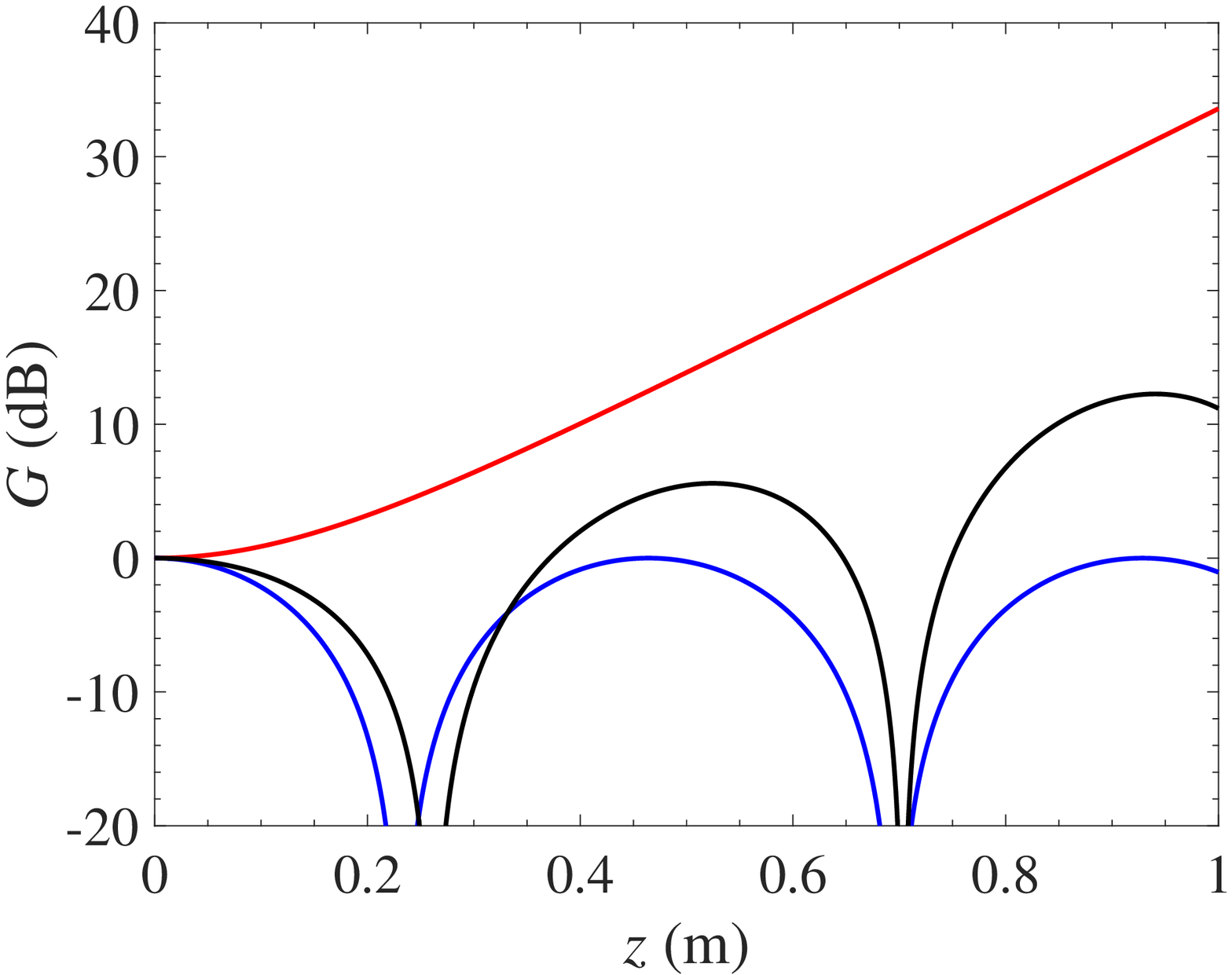}
 \caption{\label{fig:length_upper_idler} Signal power gain against amplifier length for signal frequency $f=7.5\,\mathrm{GHz}$. Red: keeping the primary amplifier terms; Blue: keeping only the upper idler terms; Black: keeping both the primary amplifier and the upper idler terms.}
 \end{figure}

\begin{figure}[ht]
 \includegraphics[width=8.6cm]{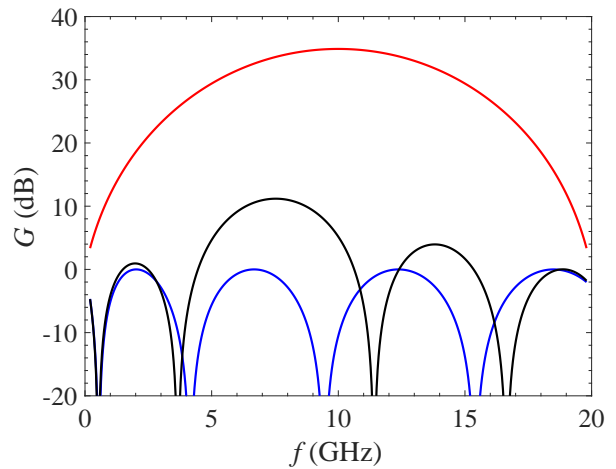}
 \caption{\label{fig:freq_upper_idler} Signal power gain against signal frequency. Red: keeping the primary amplifier terms; Blue: keeping only the upper idler terms; Black: keeping both the primary amplifier and the upper idler terms.}
 \end{figure}

Figure~\ref{fig:length_upper_idler} shows power gain plotted against amplifier length, in the presence/absence of upper idlers for a signal frequency of $f=7.5\,\mathrm{GHz}$. As seen, the addition of the upper idler terms introduces a spatial modulation into the gain. In other words, the gain does not increase monotonically with length, but can actually go to zero if the wrong length is chosen. Moreover, even when the gain is at a maximum, it is reduced significantly compared with the ideal case. The spatial modulation is due to energy being periodically siphoned away from the signal mode into the upper idler modes. It is interesting to note that the signal is attenuated for small lengths, which occurs because the signal mode always looses energy into the empty upper idler modes at the start of the amplification process. This initial attenuation has been noted by Jurkus et al. in \citenumns{Jurkus_1960}. Figure~\ref{fig:freq_upper_idler} shows power gain plotted against frequency in the presence/absence of upper idlers. Depending on the length of the TWPA, the frequency of maximum gain can shift away from the pump frequency. Again, it can be seen that the maximum achievable gain is reduced by the presence of the upper idler modes.

\section{Quantum noise}

The quantum noise resulting from the travelling-wave amplification process can be calculated by adapting the analysis developed by Caves et al. \citenumns{Caves_1982,Caves_2012} for phase-preserving linear amplifiers. The analysis shows that for a general signal $\hat{E}$ of the form
\begin{align}
  \hat{E}=\frac{1}{2}\left(\hat{a}e^{-i\omega t}+\hat{a}^\dagger e^{i\omega t}\right)\,,
\end{align}
the variance, which characterises the noise, is given by
\begin{align}
  \left\langle\left(\Delta\hat{E}\right)^2\right\rangle&=\frac{1}{2}\left\langle\left|\hat{a}\right|^2\right\rangle \\ \notag
  &=\frac{1}{2}\left(\frac{1}{2}\left\langle\left\{\hat{a},\hat{a}^\dagger\right\}\right\rangle-\left\langle\hat{a}\right\rangle\left\langle\hat{a}^\dagger\right\rangle\right)\,
\end{align}
where $\left\{\hat{a}_s,\hat{a}_s^\dagger\right\}=\hat{a}_s\hat{a}_s^\dagger+\hat{a}_s^\dagger\hat{a}_s$ is the anti-commutator.
The  analysis then proceeds by requiring $\hat{a}_{\mathrm{out}}=g\hat{a}_{\mathrm{in}}+\hat{L}^\dagger_{\mathrm{noise}}$, where $\hat{a}_{\mathrm{out}}$ is the mode operator at the output, $\hat{a}_{\mathrm{in}}$ is the mode operator at the input, $g$ is the amplitude gain, and $\hat{L}^\dagger_{\mathrm{noise}}$ is the additional noise operator. The analysis recognizes that in order for both $\hat{a}_{\mathrm{out}}$ and $\hat{a}_{\mathrm{in}}$ to obey the canonical bosonic commutation relation, $\hat{L}^\dagger_{\mathrm{noise}}$ must be constrained by $\left[\hat{L}_{\mathrm{noise}},\hat{L}^\dagger_{\mathrm{noise}}\right]=g^2-1$. This non-zero commutation relation, through the uncertainty principle, places in a lower bound on the output noise:
\begin{align}
    \left\langle\left|\hat{a}_{\mathrm{out}}\right|^2\right\rangle \geq g^2\left\langle\left|\hat{a}_{\mathrm{in}}\right|^2\right\rangle+\frac{1}{2}\left(g^2-1\right)\,.
\end{align}
When $g\gg1$, there is, \textit{at least}, half a quantum of zero-point fluctuation, which is known as the standard quantum limit of linear amplifiers.

In order to apply this analysis to TWPAs, we first modify the results to suit our definition of current given in equations~(\ref{eq:current}-\ref{eq:modes_expansion}):
\begin{align}
  \left\langle\left(\Delta\hat{I}_{s}\right)^2\right\rangle=C_{u}\left( \frac{1}{2}\left\langle\left\{\hat{a}_s,\hat{a}_s^\dagger\right\}\right\rangle-\left\langle\hat{a}_s\right\rangle\left\langle\hat{a}_s^\dagger\right\rangle \right) \,, \label{eq:Caves}
\end{align}
where $C_{u}=\hbar\omega_s/(Z_0 \tau)$ is a constant arising from unit conversion. Using the vacuum input state $\left|\mathrm{s}\right\rangle\left|\mathrm{i}\right\rangle=\left|0\right\rangle\left|0\right\rangle$, the zero-point noise is given by
\begin{align}
  \left\langle\left(\Delta\hat{I}_{s,ZP}\right)^2\right\rangle = \frac{C_u}{2}\,.
\end{align}

In the general analysis, the form of the noise operator $\hat{L}^\dagger_{\mathrm{noise}}$ is not known and can only be constrained by canonical commutation relation. In this TWPA analysis, however, we know the form of this additional term explicitly from equation~(\ref{eq:TWPA_soln_1}). Using the mode evolution derived from previous sections, again operating the TWPA with input state $\left|s\right\rangle\left|i\right\rangle=\left|a_{s,0}\right\rangle\left|0\right\rangle$, we get
\begin{align}
  \left\langle\left(\Delta\hat{I}_{s}\right)^2\right\rangle = C_{u}\left[g^2\left(\frac{1}{2}\left\langle\left\{\hat{a}_{s,0},\hat{a}_{s,0}^\dagger\right\}\right\rangle-\left\langle\hat{a}_{s,0}\right\rangle\left\langle\hat{a}_{s,0}^\dagger\right\rangle\right)\right. \notag \\
  \left.+\frac{1}{2}g'^2\left\langle\left\{\hat{a}_{i,0},\hat{a}_{i,0}^\dagger\right\}\right\rangle\right]\,,
\end{align}
where $g=\cosh(C_{i,s}z)$ is the signal amplitude gain, and $g'=\sinh(C_{i,s}z)$ is the idler amplitude gain. The last term arises from mixing the signal mode with the idler mode, and can be simplified by noting that the idler input state is the vacuum state. Thus
\begin{align}
  \left\langle\left\{\hat{a}_{i,0},\hat{a}_{i,0}^\dagger\right\}\right\rangle\ = 1\,.
\end{align}
In the large gain limit $g\approx g'$, the noise simplifies to
\begin{align}
  \left\langle\left(\Delta\hat{I}_{s}\right)^2\right\rangle = g^2\left[\left\langle\left(\Delta\hat{I}_{s,0}\right)^2\right\rangle+\left\langle\left(\Delta\hat{I}_{s,ZP}\right)^2\right\rangle\right]\,,
\end{align}
where the first term is the amplified noise pre-existing in the amplifier input and the second term is the SQL of an additional zero-point fluctuation arising from the signal mode mixing with the idler mode.

\section{Thermal noise}

Classically, the thermal noise generated by a lossy transmission line is modelled through the use of distributed noise generators \citenumns{Chang_1960,BURGESS_1962,Agouridis_1987}. In this section, we first describe the evolution of frequency modes in the presence of distributed sources, and then we relate the distributed sources to thermal noise.

First, we introduce a general distributed source term $V_{src}(z,t)$. In the presence of this source term, the momentum flux is given by
\begin{align}
  \hat{\mathcal{G}}=\frac{1}{2}\left(C\hat{V}(z,t)^2+L\hat{I}(z,t)^2\right)+\frac{1}{v}V_{src}(z,t)\hat{I}(z,t)\,. \label{eq:Source_momentum}
\end{align}

Integrating over time, we obtain a total momentum operator of the following form
\begin{align}
  \hat{G}=\hat{G}_0+\hat{G}_{4WM}+\hat{G}_S\,,
\end{align}
where
\begin{align}
  \hat{G}_0=\int \frac{d\omega}{2\pi}\frac{\hbar\omega}{v}\hat{d}^\dagger(\omega,z)\hat{d}(\omega,z)
\end{align}
is the linear transmission line term,
\begin{align}
  \hat{G}_{4WM}=&-\frac{1}{16}\left(I_{p,0}/I_*\right)^2\frac{\hbar}{v}\int \frac{d\omega_s}{2\pi}\sqrt{\omega_s\omega_i} \notag \\
  &\left[e^{i2\Phi_p}\hat{c}^\dagger(\omega_s,z)\hat{c}^\dagger(\omega_i,z)+e^{-i2\Phi_p}\hat{c}(\omega_s,z)\hat{c}(\omega_i,z)\right]
\end{align}
is the parametric amplifier term, and
\begin{align}
  \hat{G}_S=&-\frac{i}{v}\sqrt{\frac{1}{2Z_0}}\notag \\
  &\int_{t_0}^{t_0+\tau}dt V_{src}(z,t) \int_0^{\infty}\frac{d\omega}{2\pi}\sqrt{\hbar\omega}\left[\hat{d}(\omega,z)-\hat{d}^\dagger(\omega,z)\right]
\end{align}
is the source momentum term.

The momentum operators can then be evaluated to yield the following equations of motion
\begin{align}
  \frac{d}{dz}\hat{c}(\omega_s,z)=N_{src}(\omega_s,z)-iC_{\mathrm{NL}}(\omega_s)e^{i2\Phi_p}\hat{c}^\dagger(\omega_i,z) \\
  \frac{d}{dz}\hat{c}^\dagger(\omega_i,z) =\bar{N}_{src}(\omega_i,z)+iC_{\mathrm{NL}}(\omega_i)e^{-i2\Phi_p}\hat{c}(\omega_s,z)\,,
\end{align}
where
\begin{align}
  N_{src}(\omega_s,z)=&\frac{-1}{v\hbar}\sqrt{\frac{1}{2Z_0}}V_{src}(\omega_s,z)e^{-ikz}\,,\\
  V_{src}(\omega_s,z)=&\int_{t_0}^{t_0+\tau} dt V_{src}(z,t)\sqrt{\hbar\omega_s}e^{i\omega_s t}\,, \\
  C_{\mathrm{NL}}(\omega_s)=&C_{\mathrm{NL}}(\omega_i) \notag \\
  =&\frac{1}{8v}\left(I_{p,0}/I_*\right)^2\sqrt{\omega_s\omega_i}\,, = C_{\mathrm{NL}}
\end{align}
and $\bar{N}_{src}$ denotes the complex conjugate of $N_{src}$.

It can be shown that the above system of equations has solution
\begin{align}
  \hat{c}(\omega_s,z)=&\left[\hat{c}_0(\omega_s)\cosh(C_{\mathrm{NL}}z)-ie^{i2\Phi_p}\hat{c}_0^\dagger(\omega_i)\sinh(C_{\mathrm{NL}}z)\right]\notag \\
  &+\int_0^z dz' \cosh(C_{\mathrm{NL}}(z-z'))N_{src}(\omega_s,z')\notag \\
  &-\int_0^z dz' ie^{i2\Phi_p}\sinh(C_{\mathrm{NL}}(z-z'))\bar{N}_{src}(\omega_i,z')\,, \label{eq:Thermal_Solution}
\end{align}
where $\hat{c}_0(\omega_s)$ is the signal mode operator at the input of the amplifier.

The first term of equation~(\ref{eq:Thermal_Solution}) corresponds to the amplification of the input signal through the usual parametric process in the absence of distributed sources; the second term is the amplification of `signal' generated by the distributed sources along the amplifier; and the third term is mixing of idler frequency modes generated by the distributed sources into the signal frequency through parametric frequency mixing.

At this stage we have not yet constrained the distributed source term $N_{src}(\omega,z)$ to any particular physical process. In order relate $N_{src}(\omega,z)$ to thermal noise, we compare our resultant noise spectra from a TWPA without input signal to the well known Nyquist-Johnson result for a small section of transmission line with length $\Delta z$, microwave resistance per unit length $R$, and temperature $T$:
\begin{align}
  \frac{d\left\langle\hat{V}^2(\omega,\Delta z)\right\rangle}{d\omega}=&\frac{Z_0}{2\pi}\hbar\omega\int_0^{\Delta z} dz'\int_0^{\Delta z} dz'' \notag \\ &\left\langle\bar{N}_{src}(\omega,z')N_{src}(\omega,z'')\right\rangle \\
  =&4\pi\hbar\omega\coth\left(\frac{\hbar\omega}{2k_B T}\right)R\Delta z\,. 
\end{align} Here $\Delta z$ is chosen such that it is much shorter than all length scales arising from nonlinearity mixing. The quantum symmetrized generalization of the Nyquist-Johnson result, i.e. the Callen-Welton formula \citenumns{Callen_1951, Ginzburg_1987}, has been used here. We have also made the assumption that the attenuation caused by $R$ is small compared with the TWPA gain. Simplifying, we get the following spatial correlation function for the distributed thermal noise sources \citenumns{Rytov_1989}
\begin{align}
  \left\langle\bar{N}_{src}(\omega,z)N_{src}(\omega,z')\right\rangle = M(\omega)\delta(z-z')&\,\,\,\,\mathrm{where}\\
  M(\omega)=2(2\pi)^2\coth\left(\frac{\hbar\omega}{2k_B T}\right)\frac{R}{Z_0}&\,.\label{eq:Thermal}
\end{align}

From the correlation of the noise sources, we calculate the noise spectra $S$ of the full TWPA to be
\begin{align}
  S =& \frac{d\left\langle\hat{V}^2(\omega_s,z)\right\rangle}{d\omega_s} \notag \\
  =& \frac{Z_0}{2\pi}\hbar\omega_s\int_0^z dz' \notag \\
  &\left[M(\omega_s)\cosh^2(C_{\mathrm{NL}}(z-z'))\right.\, \notag \\
  &\left.\,+M(\omega_i)\sinh^2(C_{\mathrm{NL}}(z-z'))\right]\,
\end{align}
The first term, involving $\omega_s$, is the thermal noise at the signal frequency amplified by propagation along the line; the second, term involving $\omega_i$, is the thermal noise generated at the idler frequency mixed into the signal frequency. This second term is unique to amplifiers based on frequency-mixing.

This result is intuitive in both the low gain and high gain limits. In the low gain limit where the nonlinearity of the medium is negligible,
\begin{align}
  \cosh^2(C_{\mathrm{NL}}(z-x))\rightarrow1\,,  \\
  \sinh^2(C_{\mathrm{NL}}(z-x))\rightarrow0\,,
\end{align}
and the noise spectrum is same as the spectrum of a linear transmission line:
\begin{align}
  \frac{d\left\langle\hat{V}^2(\omega_s,z)\right\rangle}{d\omega_s} = 4\pi\hbar\omega\coth\left(\frac{\hbar\omega}{2k_B T}\right)Rz\,.
\end{align}

In the high gain limit where the nonlinearity results in significant amplification,
\begin{align}
  \cosh^2(C_{\mathrm{NL}}(z-x))&\rightarrow \frac{1}{4}\exp(2C_{\mathrm{NL}}(z-x)) \gg1\,,  \\
  \sinh^2(C_{\mathrm{NL}}(z-x))&\rightarrow \frac{1}{4}\exp(2C_{\mathrm{NL}}(z-x))\,,
\end{align}
the noise spectrum involves a symmetrical contribution from the signal and idler frequencies:
\begin{align}
  \frac{d\left\langle\hat{V}^2(\omega_s,z)\right\rangle}{d\omega_s} = \frac{Z_0}{2\pi}\hbar\omega_s\frac{e^{2C_{\mathrm{NL}}z}}{8C_{\mathrm{NL}}}\left[M(\omega_s)+M(\omega_i)\right]\,.
\end{align}

\begin{figure}[ht]
\includegraphics[width=8.6cm]{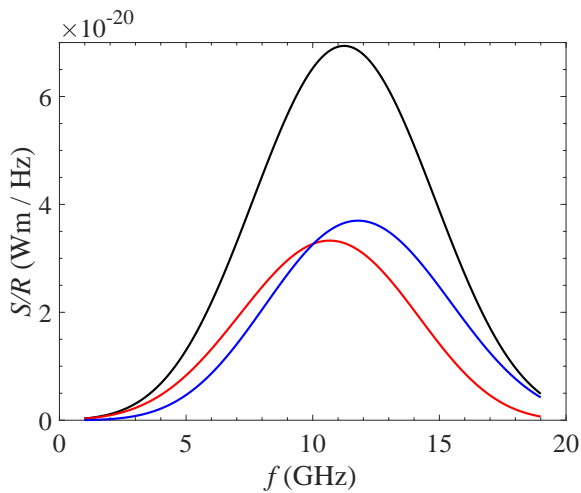}
\caption{\label{fig:thermal_noise} Noise spectra of a parametric amplifier normalized by its per-unit-length resistance $R$ at temperature $T=0.2\,\mathrm{K}$. Black line: plot of total thermal noise against frequency; red line: plot of thermal noise against frequency arising from signal frequency; blue line: plot of thermal noise against frequency arising from idler frequency via mixing.}
\end{figure}

The analysis method described in this section is applicable to general TWPAs without being restricted to a specific transmission line geometry or material. Here, however, we perform a representative calculation for a coplanar waveguide fabricated using titanium. The amplifier has inductance per unit length $L_0=8.1\times10^{-7}\,\mathrm{Hm^{-1}}$, capacitance per unit length $C=2.8\times10^{-10}\,\mathrm{Fm^{-1}}$, length $L_{PA}=1\,\mathrm{m}$, and pump frequency $f_{p}=10\,\mathrm{GHz}$. The amplifier is operated at $I_{p,0}/I_*=1/5$, and temperature $T=0.2\,\mathrm{K}$. Figure~\ref{fig:thermal_noise} shows thermal noise plotted against frequency. The thermal noise is normalized against the material-dependent resistance per unit length $R$ which is non-zero at microwave frequencies even for a superconducting transmission line. The black line shows the total thermal noise, whilst the red and blue lines show the contributions from the signal and mixed-in idler frequencies respectively. As seen, the total noise can be significantly higher than the noise from the signal frequency only, especially near the pump frequency.

Naively, one might want to operate TWPAs at higher signal frequencies in order to reduce thermal noise. However, as the signal frequency increases, the idler frequency decreases and contributes more significantly to the total thermal noise. In many TWPA applications, the experimenter may wish to operate in the regime where $\omega_s\approx\omega_p\approx\omega_i$, in order to maximise amplifier gain. In this regime, the frequency-mixing doubles the amount of thermal noise.

\begin{figure}[ht]
\includegraphics[width=8.6cm]{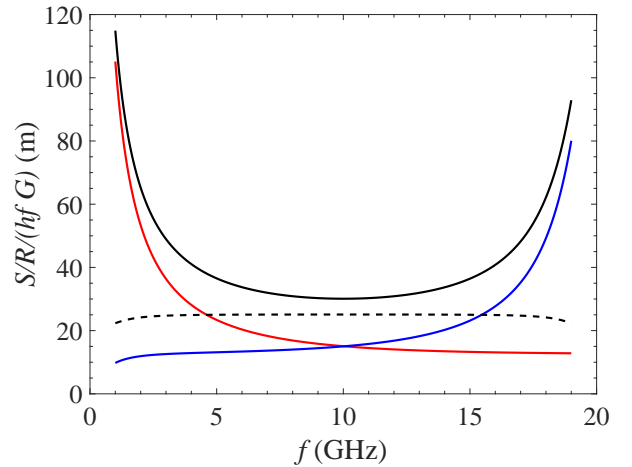}
\caption{\label{fig:thermal_noise_normalized} Noise spectra of a parametric amplifier normalized by its per-unit-length resistance $R$, amplifier power gain $G$, and quantum energy from signal mode $hf=\hbar\omega$. Solid black line: plot of total thermal noise against frequency at temperature $T=0.2\,\mathrm{K}$; solid red line: plot of thermal noise against frequency arising from the signal frequency at temperature $T=0.2\,\mathrm{K}$; solid blue line: plot of thermal noise against frequency arising from the idler frequency via frequency mixing at temperature $T=0.2\,\mathrm{K}$; dashed black line: plot of total thermal noise against frequency at temperature $T=0.01\,\mathrm{K}$.}
\end{figure}

Figure~\ref{fig:thermal_noise_normalized} shows the thermal noise from the same amplifier further normalized against amplifier power gain $G$ and quantum energy $hf=\hbar\omega$. This normalization allows thermal noise to be measured in terms of noise added to the input. (To fully express thermal noise in terms of number of input noise quanta, one further requires information about the readout bandwidth as well as the resistance per unit length $R$ of the amplifier material.) This metric also emphasizes the importance of the idler contribution when the signal frequency is higher than the pump frequency: at around $15\,\mathrm{GHz}$, the contribution from the idler is almost double compared to that from the signal. In terms of absolute magnitudes, the thermal noise can be comparable to the quantum noise depending on the choice of readout bandwidth and material resistivity. TWPAs are often operated at pump power levels that are close to the onset of significant resistivity (see supplement of \citenumns{Eom_2012}). When choosing the optimum pump power level, it is important to ensure that the growth of noise does not outstrip the growth in gain since both thermal noise and amplifier gain are high at high pump levels. The dotted black line in figure~\ref{fig:thermal_noise_normalized} shows the total normalized thermal noise at $T=0.01\,\mathrm{K}$. At signal frequencies close to the pump frequency, where the gain is high, the noise is in the quantum limit since $k_{B}T/\hbar\omega_{p}=0.02\ll1$. The quantum thermal noise here arises from quantum fluctuations of modes associated with the non-zero microwave resistivity of the amplifier superconducting line. It is distinct from the intrinsic quantum noise which gives rise to the SQL and is a result of the wave mixing between the signal and the idler. One of the advantages of TWPAs is that their operational temperature is set by the critical temperature of the superconducting material. As a result, they can be operated at very high temperatures, even up to $4\,\mathrm{K}$ \citenumns{Adamyan_2016}. Here we see that the ratio $k_{B}T$ to $\hbar\omega_{p}$ determines the magnitude of overall thermal noise through equation~(\ref{eq:Thermal}). This can place additional restrictions on the useful temperature range within which a TWPA can produce desirable level of noise performance.


\section{Squeezing amplifier using three-wave mixing}

Although TWPAs can operate with noise temperatures approaching the SQL, we are not aware of any superconducting TWPA being used for the generation and amplification of squeezed states. In this and the next section, we discuss whether generation and amplification are in principle possible. In order to use a TWPA to generate a squeezed state, it must be operated in the degenerate amplifier mode where $\omega_s=\omega_i$. This poses a challenge for four-wave mixing based TWPAs because the signal frequency is now indistinguishable from the pump frequency $\omega_s=\omega_p$. This problem can be overcome through the introduction of a strong DC bias current $I_{DC}$, thereby operating the TWPA in an effectively three-wave mixing mode.

In the presence of DC bias, the distributed inductance becomes
\begin{align}
  L=L_0\left[1+\left(\frac{I_{DC}}{I_*}\right)^2+2\left(\frac{I_{DC}I}{I_*^2}\right)+\left(\frac{I}{I_*}\right)^2\right]\,,
\end{align}
where $I$ contains all of the AC terms: including pump, signal, and idler. At this stage we do not operate the amplifier
as a degenerate amplifier, and so distinguish between the signal and idler. In the case where $I_{DC}\gg I_{p,0}$, the last term of the above equation, which is responsible for four-wave mixing, can be treated as approximately zero. The total momentum operator is then given by
\begin{align}
  \hat{G}=& \hat{G}_0+ \hat{G}_{3WM}\,,\\
  \hat{G}_{3WM} =&- \hbar D_{i,s}\left[\exp(i\Phi_p)\hat{a}_i^\dagger\hat{a}_s^\dagger\right.\,\notag \\
  &\left.+\exp(-i\Phi_p)\hat{a}_i\hat{a}_s\right]\,,
\end{align}
where
\begin{align}
  \omega_i&=\omega_p-\omega_s\,,\\
  D_{i,s}&=\frac{L_0}{2L_1}\left(\frac{I_{DC}I}{I_*^2}\right)\frac{\sqrt{\omega_i\omega_s}}{v}\,, \\
  L_1&=L_0\left[1+\left(\frac{I_{DC}}{I_*}\right)^2\right]\,.
\end{align}

The form of $\hat{G}_{3WM}$ is similar to that of $\hat{G}_{4WM}$ except for certain redefinitions of constants, most notably $2\omega_p \rightarrow \omega_p$, which shifts the three-wave mixing gain pattern to be centred around $\omega_p/2$ instead of $\omega_p$. In the degenerate case where $\omega_s=\omega_i=\omega_p/2$, the equations of motion are given by
\begin{align}
  \hat{a}_s=&\hat{a}_{s,0}\cosh(2D_{i,s}z)\notag \\
  &-i\exp(i\Phi_p)\hat{a}_{s,0}^\dagger\sinh(2D_{i,s}z) \, \\
  \hat{a}_s^\dagger=&\hat{a}_{s,0}^\dagger\cosh(2D_{i,s}z) \notag \\
  &+i\exp(-i\Phi_p)\hat{a}_{s,0}\sinh(2D_{i,s}z) \, ,
\end{align}
where $\hat{a}_{s,0}$ is the signal mode at the input of the amplifier.

We define two quadrature envelope functions \citenumns{gerry_knight_2004,Octavian_2013} by
\begin{align}
  \hat{X}_s(\Phi_p,z)=\hat{a}_s+\hat{a}_s^\dagger \\
  \hat{Y}_s(\Phi_p,z)=\hat{a}_s-\hat{a}_s^\dagger \,.
\end{align}
When $\Phi_p=\pi/2$,
\begin{align}
  \hat{X}_s\left(\frac{\pi}{2},z\right)=e^{2D_{s,s}z}X_s\left(\frac{\pi}{2},0\right) \\
  \hat{Y}_s\left(\frac{\pi}{2},z\right)=e^{-2D_{s,s}z}Y_s\left(\frac{\pi}{2},0\right) \,.
\end{align}
In other words, one quadrature is amplified, and the other is attenuated; the degenerate TWPA is thus functioning as a squeezing amplifier. It seems that a TWPA can in principle be used to generate broadband squeezed states.

\section{Amplifying pre-squeezed signal}

Another potential application of a TWPA is to place it \textit{after} a squeezing amplifier. In this section we evaluate the characteristics of a non-degenerate TWPA when a squeezed state is applied to its input. The initial state is given by $\left|s\right\rangle\left|i\right\rangle=\left|a_{s,0}\right\rangle\left|0\right\rangle$, where the idler refers to the TWPA idler, i.e. $\omega_i=2\omega_p-\omega_s$. For simplicity and consistency of notation, we assume that the squeezing amplifier ahead of the TWPA operates the same way as described in the previous section. We define $L_{sq}$ and $\Phi_{p,sq}$ as the length and the pump phase of the squeezing amplifier, and $L_{PA}$ and $\Phi_{p,PA}$ as the length and the pump phase of the TWPA. Subscript $0$ indicates the input to the squeezing amplifier, subscript $in$ indicates the input to the TWPA, and subscript $out$ indicates the output of the TWPA.

After the squeezing amplifier, the signal mode is given by
\begin{align}
  \hat{a}_{s,in}=\hat{a}(L_{sq})_s=&\hat{a}_{s,0}\cosh(2D_{i,s}L_{sq})\notag \\
  &-i\exp(i\Phi_{p,sq})\hat{a}_{s,0}^\dagger\sinh(2D_{i,s}L_{sq}) \, \\
  \hat{a}_{s,in}^\dagger=\hat{a}(L_{sq})_s^\dagger=&\hat{a}_{s,0}^\dagger\cosh(2D_{i,s}L_{sq}) \notag \\
  &+i\exp(-i\Phi_{p,sq})\hat{a}_{s,0}\sinh(2D_{i,s}L_{sq}) \, .
\end{align}

Using the above as the input to the TWPA, i.e. equations~(\ref{eq:TWPA_soln_1}) and (\ref{eq:TWPA_soln_2}), we obtain
\begin{align}
  \hat{a}_{s,out}=&\hat{a}_{s,0}\cosh(2D_{i,s}L_{sq})\cosh(C_{i,s}L_{PA})\notag \\
  &-i\exp(i\Phi_{p,sq})\hat{a}_{s,0}^\dagger\sinh(2D_{i,s}L_{sq})\cosh(C_{i,s}L_{PA}) \notag \\
  &-i\exp(i2\Phi_{p,PA})\hat{a}_{i,in}^\dagger\sinh(C_{i,s}L_{PA})\, \\
  \hat{a}_{s,out}^\dagger=&\hat{a}_{s,0}^\dagger\cosh(2D_{i,s}L_{sq})\cosh(C_{i,s}L_{PA}) \notag \\
  &+i\exp(-i\Phi_{p,sq})\hat{a}_{s,0}\sinh(2D_{i,s}L_{sq})\cosh(C_{i,s}L_{PA}) \notag \\
  &+i\exp(-i2\Phi_{p,PA})\hat{a}_{i,in}\sinh(C_{i,s}L_{PA})\, .
\end{align}

When $\Phi_p=\pi/2$,
\begin{align}
  \left\langle \hat{X}_{s,out}\right\rangle=g_{PA}g_{sq}\left\langle X_{s,0}\right\rangle\,\\
  \left\langle \hat{Y}_{s,out}\right\rangle=\frac{g_{PA}}{g_{sq}}\left\langle Y_{s,0}\right\rangle\,,
\end{align}
where $g_{PA}=\cosh(C_{i,s}L_{PA})$ is the gain of the TWPA, and $g_{sq}=\exp(2D_{i,s}L_{sq})$ is the gain of the squeezing amplifier. Hence we see that the overall gain of the two-stage amplification process is cascaded in the way that would be expected.

When calculating the uncertainties in the two quadratures at the output, to find the noise, we must be careful not to apply Caves's formula, equation~(\ref{eq:Caves}), directly, because the assumption that the noise is phase-insensitive is not true in the case of squeezing amplifiers. Instead, we find the noise by using
\begin{align}
  \langle \Delta \hat{X}_{s,out}^2\rangle =& \langle \hat{X}_{s,out}^2\rangle - \langle \hat{X}_{s,out}\rangle ^2 \\
  =&\langle\hat{a}_{s,out}\hat{a}_{s,out}+\hat{a}_{s,out}^\dagger\hat{a}_{s,out}^\dagger\rangle \notag \\
  -&\langle\hat{a}_{s,out}\rangle\langle\hat{a}_{s,out}\rangle \notag \\
  -&\langle\hat{a}_{s,out}^\dagger\rangle\langle\hat{a}_{s,out}^\dagger\rangle \notag \\
  +&\langle\{\hat{a}_{s,out},\hat{a}_{s,out}^\dagger\}\rangle \\
  -&2\langle\hat{a}_{s,out}\rangle\langle\hat{a}_{s,out}^\dagger\rangle\,.
\end{align}

In the large-TWPA-gain limit, $g_{PA}=\cosh(C_{i,s}L_{PA})\approx\sinh(C_{i,s}L_{PA})$, and the noise is given by
\begin{align}
  \langle \Delta \hat{X}_{s,out}^2\rangle =g_{PA}^2\left[g_{sq}^2\langle \Delta \hat{X}_{s,0}^2\rangle +\langle \Delta \hat{X}_{s,ZP}^2\rangle\right]\,, \\
  \langle \Delta \hat{Y}_{s,out}^2\rangle =g_{PA}^2\left[g_{sq}^{-2}\langle \Delta \hat{Y}_{s,0}^2\rangle +\langle \Delta \hat{Y}_{s,ZP}^2\rangle\right]\,,
\end{align}
where $\langle \Delta \hat{X}_{s,ZP}^2\rangle$ and $\langle \Delta \hat{Y}_{s,ZP}^2\rangle$ are the zero-point fluctuations of the $X$ and $Y$ quadratures respectively.

The total noise thus has two contributions, as before: the first term arises from cascading the gains of the squeezing and post-squeezed amplification stages; the second term arises from the zero-point noise of the TWPAs idler . This observation has important implications if the squeezing amplifier is being used to create a squeezed vacuum state. The TWPA does not maintain or enhance the imbalance between the two signal quadratures. Instead, it diminishes the imbalance through the introduction of the idler's zero-point noise. This scenario has experimental relevance in the case where a TWPA is used to monitor the output of a squeezing amplifier.

\section{Conclusion}

Superconducting TWPAs having noise temperatures approaching the standard quantum limit have been demonstrated by a number of investigators. These successes motivate the need to better understand certain aspects of behaviour. In this paper we have described theoretical analyses, based on a quantised momentum-flux approach, and numerical simulations of the following: (i) the effect of having impedance discontinuities at the terminations; (ii) the role of upper idler modes on the gain profile; (iii) the modal origins of quantum and thermal noise; (iv) the generation of squeezed states; and (v) the preservation of pre-squeezed states during amplification.

In the case of (i), it seems that in the absence of a reflected pump, a backward propagating signal suffers no attenuation beyond that ordinarily associated with ohmic and dielectric loss. In the presence of a reflected pump, a backward propagating signal can be amplified. The lack of attenuation in the reverse direction can have a marked effect on the gain profile, and in the extreme case may cause instability. Our simulations show that mismatch places a restriction on the maximum pump power, and maximum forward gain, beyond which the loop gain exceeds unity. 
In the case of (ii), upper idler modes can siphon energy away from the signal modes, resulting in suppressed gain. The signal gain varies periodically with propagation distance, meaning that one cannot assume increasing the length of an amplifier will increase its gain. (iii) We have analysed the noise associated with the signal mode to show that an ideal TWPA is indeed quantum-noise limited. Additionally, we have shown that in the presence of internally generated thermal noise, frequency mixing between the signal and idler modes significantly modifies the propagating noise spectrum. A fundamental observation is that when the signal frequency is close to the pump frequency, there is a near doubling of the thermal noise as compared with that generated by the losses in the signal mode alone. To understand (iv), we studied the propagation of the quadrature components of a signal in a superconducting TWPA, and have shown that highly squeezed states can, in principle at least, be generated. However in the context of (v), care is needed when amplifying a pre-squeezed state because the squeezing advantage is generally reduced.

\appendix

\section{Quantum telegrapher's equations}
When applying an analysis framework based on quantum mechanics, it is useful to demonstrate that classical results can be derived from the more general quantum approach. Here we show that, in the absence of nonlinearity, the quantum equations of motion are analogues to the classical telegrapher's equations.

The total momentum operator is given by
\begin{align}
\hat{G}&=\int_{t_0}^{t_0+\tau}dt\,\frac{1}{2}\left(C\hat{V}(z,t)^2+L\hat{I}(z,t)^2\right) \,, \\
&= \int_{t_0}^{t_0+\tau}dt\,\frac{1}{2}\left(C\left(\frac{\partial\hat{\phi}(z,t)}{\partial t}\right)^2+L\hat{I}(z,t)^2\right)
\end{align}
where
\begin{align}
  \hat{I}(z,t) = -\frac{1}{L}\frac{\partial\hat{\phi}}{\partial z}\,, \\
  \hat{V}(z,t) = \frac{\partial\hat{\phi}}{\partial t}\,,
\end{align}
and $\hat{\phi}$ is the generalized flux operator.

The relevant equal position commutation relation is given by
\begin{align}
  \left[\hat{\phi}(z,t),\hat{I}(z,t')\right]=-i\hbar\delta(t-t') \,.
\end{align}

The momentum picture equation of motion, i.e.
\begin{align}
  i\hbar \frac{d\hat{A}}{dz}=\left[\hat{A},\hat{G}\right] \,.
\end{align}
can then be applied to operators $\hat{\phi}(z,t)$ and $\hat{I}(z,t)$.

Equation of motion for $\hat{\phi}(z,t)$ yields
\begin{align}
  -L\frac{\partial}{\partial t}\hat{I}(z,t) &= \frac{\partial}{\partial z}\frac{\partial}{\partial t}\hat{\phi}(z,t) \\
  &= \frac{\partial}{\partial z}\hat{V}(z,t)\,.
\end{align}

Equation of motion for $\hat{I}(z,t)$ yields
\begin{align}
  \frac{\partial}{\partial z}\hat{I}(z,t) &= -C\frac{\partial}{\partial t}\frac{\partial}{\partial t}\hat{\phi}(z,t) \\
  &= -C\frac{\partial}{\partial t}\hat{V}(z,t)\,.
\end{align}

Hence the classical telegrapher's equations can be obtained readily from our quantum analysis framework.

\bibliographystyle{h-physrev}
\bibliography{library}

\begin{thebibliography}{10}

\bibitem{Chaudhuri_2015}
S.~Chaudhuri, J.~Gao, and K.~Irwin,
\newblock IEEE T. Appl. Supercon. {\bf 25}, 1 (2015).

\bibitem{Erickson_2017}
R.~P. Erickson and D.~P. Pappas,
\newblock Phys. Rev. B {\bf 95}, 104506 (2017).

\bibitem{Songyuan_2019_paramp}
S.~Zhao, S.~Withington, D.~J. Goldie, and C.~N. Thomas,
\newblock Journal of Physics D: Applied Physics {\bf 52}, 415301 (2019).

\bibitem{songyuan2019_nonlinear}
S.~Zhao, S.~Withington, D.~J. Goldie, and C.~N. Thomas,
\newblock Journal of Low Temperature Physics {\bf 199}, 34 (2020).

\bibitem{Eom_2012}
B.~H. Eom, P.~K. Day, H.~G. LeDuc, and J.~Zmuidzinas,
\newblock Nat. Phys. {\bf 8}, 623 (2012).

\bibitem{Shan_2016}
W.~Shan, Y.~Sekimoto, and T.~Noguchi,
\newblock IEEE T. Appl. Supercon. {\bf 26}, 1 (2016).

\bibitem{Adamyan_2016}
A.~A. Adamyan, S.~E. de~Graaf, S.~E. Kubatkin, and A.~V. Danilov,
\newblock Journal of Applied Physics {\bf 119}, 083901 (2016),
  https://doi.org/10.1063/1.4942362.

\bibitem{Chaudhuri_2017}
S.~Chaudhuri {\em et~al.},
\newblock Appl. Phys. Lett. {\bf 110}, 152601 (2017),
  https://doi.org/10.1063/1.4980102.

\bibitem{shu2021nonlinearity}
S.~Shu {\em et~al.},
\newblock Nonlinearity and wideband parametric amplification in an nbtin
  microstrip transmission line, 2021, 2103.00656.

\bibitem{Bockstiegel_2014}
C.~Bockstiegel {\em et~al.},
\newblock Journal of Low Temperature Physics {\bf 176}, 476 (2014).

\bibitem{Vissers_2016}
M.~R. Vissers {\em et~al.},
\newblock Applied Physics Letters {\bf 108}, 012601 (2016),
  https://doi.org/10.1063/1.4937922.

\bibitem{Ranzani_2018}
L.~Ranzani {\em et~al.},
\newblock Applied Physics Letters {\bf 113}, 242602 (2018),
  https://doi.org/10.1063/1.5063252.

\bibitem{Zobrist_2019}
N.~Zobrist {\em et~al.},
\newblock Applied Physics Letters {\bf 115}, 042601 (2019),
  https://doi.org/10.1063/1.5098469.

\bibitem{Zorin_2019}
A.~Zorin,
\newblock Phys. Rev. Applied {\bf 12}, 044051 (2019).

\bibitem{Vissers_2020}
M.~R. Vissers {\em et~al.},
\newblock {Demonstration of a microwave SQUID multiplexer with
  pre-amplification from a kinetic inductance traveling-wave parametric
  amplifier},
\newblock in {\em Millimeter, Submillimeter, and Far-Infrared Detectors and
  Instrumentation for Astronomy X}, edited by J.~Zmuidzinas and J.-R. Gao Vol.
  11453, International Society for Optics and Photonics, SPIE, 2020.

\bibitem{Barends_2014}
R.~Barends {\em et~al.},
\newblock Nature {\bf 508}, 500 (2014).

\bibitem{Vijay_2012}
R.~Vijay {\em et~al.},
\newblock Nature {\bf 490}, 77 (2012).

\bibitem{Riste_2017}
D.~Rist{\`e} {\em et~al.},
\newblock npj Quantum Information {\bf 3}, 16 (2017).

\bibitem{Monfardini_2010}
{Monfardini, A.} {\em et~al.},
\newblock A\&A {\bf 521}, A29 (2010).

\bibitem{Maloney_2010}
P.~R. Maloney {\em et~al.},
\newblock {MUSIC for sub/millimeter astrophysics},
\newblock in {\em Millimeter, Submillimeter, and Far-Infrared Detectors and
  Instrumentation for Astronomy V}, edited by W.~S. Holland and J.~Zmuidzinas
  Vol. 7741, pp. 124 -- 134, International Society for Optics and Photonics,
  SPIE, 2010.

\bibitem{Endo_2012}
A.~Endo {\em et~al.},
\newblock Journal of Low Temperature Physics {\bf 167}, 341 (2012).

\bibitem{Mazin_2013}
B.~A. Mazin {\em et~al.},
\newblock Publications of the Astronomical Society of the Pacific {\bf 125},
  1348 (2013).

\bibitem{Battistelli2015}
E.~S. Battistelli {\em et~al.},
\newblock The European Physical Journal C {\bf 75}, 353 (2015).

\bibitem{Cardani_2015}
L.~Cardani {\em et~al.},
\newblock Applied Physics Letters {\bf 107}, 093508 (2015),
  https://doi.org/10.1063/1.4929977.

\bibitem{Project8_2009}
B.~Monreal and J.~A. Formaggio,
\newblock Phys. Rev. D {\bf 80}, 051301 (2009).

\bibitem{Hao_2019}
L.~Hao,
\newblock Quantum technology for cyclotron radiation detection,
\newblock Absolute Neutrino Mass Workshop at UCL, 2019.

\bibitem{Saakyan_2020}
R.~Saakyan,
\newblock Determination of neutrino mass with quantum technologies,
\newblock UK HEP Forum 2020: Quantum leaps to the dark side at Durham
  University, 2020.

\bibitem{Golwala2008}
S.~Golwala {\em et~al.},
\newblock Journal of Low Temperature Physics {\bf 151}, 550 (2008).

\bibitem{Cornell_2018}
B.~D. Cornell,
\newblock {\em A Dark Matter Search Using the Final SuperCDMS Soudan Dataset
  and the Development of a Large-Format, Highly-Multiplexed,
  Athermal-Phonon-Mediated Particle Detector},
\newblock PhD thesis, California Institute of Technology, 2018.

\bibitem{Caves_1982}
C.~M. Caves,
\newblock Phys. Rev. D {\bf 26}, 1817 (1982).

\bibitem{Renger_2020}
M.~Renger {\em et~al.},
\newblock Beyond the standard quantum limit of parametric amplification, 2020,
  2011.00914.

\bibitem{Johnson_1928}
J.~B. Johnson,
\newblock Phys. Rev. {\bf 32}, 97 (1928).

\bibitem{Giordmaine_OPA_1965}
J.~A. Giordmaine and R.~C. Miller,
\newblock Phys. Rev. Lett. {\bf 14}, 973 (1965).

\bibitem{Manley_Rowe_1956}
J.~M. {Manley} and H.~E. {Rowe},
\newblock Proceedings of the IRE {\bf 44}, 904 (1956).

\bibitem{Hansryd_2002}
J.~{Hansryd}, P.~A. {Andrekson}, M.~{Westlund}, {Jie Li}, and P.~. {Hedekvist},
\newblock IEEE Journal of Selected Topics in Quantum Electronics {\bf 8}, 506
  (2002).

\bibitem{Kylemark_2004}
P.~{Kylemark}, P.~O. {Hedekvist}, H.~{Sunnerud}, M.~{Karlsson}, and P.~A.
  {Andrekson},
\newblock Journal of Lightwave Technology {\bf 22}, 409 (2004).

\bibitem{Louisell_1961}
W.~H. Louisell, A.~Yariv, and A.~E. Siegman,
\newblock Phys. Rev. {\bf 124}, 1646 (1961).

\bibitem{Clerk_2010}
A.~A. Clerk, M.~H. Devoret, S.~M. Girvin, F.~Marquardt, and R.~J. Schoelkopf,
\newblock Rev. Mod. Phys. {\bf 82}, 1155 (2010).

\bibitem{Abram_1987}
I.~Abram,
\newblock Phys. Rev. A {\bf 35}, 4661 (1987).

\bibitem{Hutter_1990}
B.~Huttner, S.~Serulnik, and Y.~Ben-Aryeh,
\newblock Phys. Rev. A {\bf 42}, 5594 (1990).

\bibitem{Mussot_2004}
A.~Mussot {\em et~al.},
\newblock IEEE Photonics Technology Letters {\bf 16}, 1289 (2004).

\bibitem{Carter_2018}
B.~Carter and R.~Mancini,
\newblock Chapter 6 - feedback and stability theory,
\newblock in {\em Op Amps for Everyone (Fifth Edition)}, edited by B.~Carter
  and R.~Mancini, pp. 59--75, Newnes, , fifth edition ed., 2018.

\bibitem{Haidekker_2020}
M.~A. Haidekker,
\newblock 13 - robustness of feedback control systems,
\newblock in {\em Linear Feedback Controls (Second Edition)}, edited by M.~A.
  Haidekker, pp. 203--218, Elsevier, , second edition ed., 2020.

\bibitem{Jurkus_1960}
A.~{Jurkus} and P.~N. {Robson},
\newblock Proceedings of the IEE - Part B: Electronic and Communication
  Engineering {\bf 107}, 119 (1960).

\bibitem{Kononov_1989}
M.~V. {Kononov} and S.~V. {Koshevaia},
\newblock Radiofizika {\bf 32}, 626 (1989).

\bibitem{Caves_2012}
C.~M. Caves, J.~Combes, Z.~Jiang, and S.~Pandey,
\newblock Phys. Rev. A {\bf 86}, 063802 (2012).

\bibitem{Chang_1960}
K.~K.~N. Chang,
\newblock Journal of Applied Physics {\bf 31}, 871 (1960),
  https://doi.org/10.1063/1.1735710.

\bibitem{BURGESS_1962}
R.~E. Burgess,
\newblock Solid-State Electronics {\bf 4}, 25  (1962).

\bibitem{Agouridis_1987}
D.~C. {Agouridis},
\newblock IEEE Transactions on Instrumentation and Measurement {\bf IM-36}, 132
  (1987).

\bibitem{Callen_1951}
H.~B. Callen and T.~A. Welton,
\newblock Phys. Rev. {\bf 83}, 34 (1951).

\bibitem{Ginzburg_1987}
V.~L. Ginzburg and L.~P. Pitaevski{\u{\i}},
\newblock Soviet Physics Uspekhi {\bf 30}, 168 (1987).

\bibitem{Rytov_1989}
S.~M. Rytov, Y.~A. Kravtsov, and V.~I. Tatarskii,
\newblock {\em Principles of Statistical Radiophysics 3} (Springer Berlin
  Heidelberg, 1989).

\bibitem{gerry_knight_2004}
C.~Gerry and P.~Knight,
\newblock {\em Introductory Quantum Optics} (Cambridge University Press, 2004).

\bibitem{Octavian_2013}
F.~D.~O. {Schackert},
\newblock {\em {A Practical Quantum-Limited Parametric Amplifier Based on the
  Josephson Ring Modulator}},
\newblock PhD thesis, Yale University, 2013.

\end{thebibliography}
\end{document}